\documentclass{llncs}

\usepackage{latexsym}
\usepackage{epsfig}
\usepackage{xspace}

\makeatletter
\newenvironment{prog}{\vspace{1.0ex}\par
\setlength{\parindent}{1.5ex}
\setlength{\parskip}{0.0ex}
\obeylines\@vobeyspaces\tt}{\vspace{1.0ex}\noindent
}
\makeatother
\newcommand{\startprog}{\begin{prog}}
\newcommand{\stopprog}{\end{prog}\noindent}
\newcommand{\code}[1]{\mbox{\tt #1}}   
\newcommand{\ccode}[1]{``\mbox{\tt #1}''}   
\newcommand{\ttbs}{\mbox{\tt\char92}}
\newcommand{\ol}[1]{\overline{#1}}  

\newcommand{\cb}{CurryBrowser\xspace}
\catcode`\_=\active
\let_=\sb
\catcode`\_=12
\makeatother

\begin{document}
\sloppy

\title{\cb: A Generic Analysis Environment for Curry Programs\thanks{
A preliminary version of this paper appeared in the
Proceedings of the International Workshop on Curry and
Functional Logic Programming (WCFLP 2005), ACM Press, pp.\ 43-48, 2005.
This work has been partially supported by the
German Research Council (DFG) under grant Ha 2457/5-1
and the NSF under grant CCR-0218224.}}
\author{Michael Hanus}
\institute{Institut f\"ur Informatik, Christian-Albrechts-Universit\"at Kiel\\
D-24098 Kiel, Germany\\
\email{mh@informatik.uni-kiel.de}
}

\maketitle

\begin{abstract}
We present \cb, a generic analysis environment for the
declarative multi-paradigm language Curry.
\cb supports browsing through the program code of an
application written in Curry, i.e., the main module and all 
directly or indirectly imported modules.
Each module can be shown in different formats (e.g., source code,
interface, intermediate code) and, inside each module,
various properties of functions defined in this module can be analyzed.
In order to support the integration of various program analyses,
\cb has a generic interface to connect local and global analyses
implemented in Curry.
\cb is completely implemented in Curry using libraries
for GUI programming and meta-programming.
\end{abstract}

\section{Overview}
\label{sec-overview}

\cb is intended as a tool to support the analysis of declarative programs.
It can be used to browse through an implementation written
in the declarative multi-paradigm language Curry \cite{Hanus97POPL,Hanus06Curry},
analyze properties of individual or all functions defined in a module,
or visualize dependencies between modules or functions.
It can be also used as a testbed for program analyses
(the analyses of functional logic programs is still ongoing research)
since it supports the easy integration of new program analyses
by a generic interface.
The implementation of \cb is based
on an intermediate language to which functional, logic,
and also integrated functional logic programs can be compiled
(e.g., see \cite{AntoyHanus00FROCOS,%
AntoyHanusMasseySteiner01PPDP,Hortala-GonzalezUllan01}).
Thus, it is also adaptable to other declarative languages.

\begin{figure*}[t]
\begin{center}
  \epsfig{file=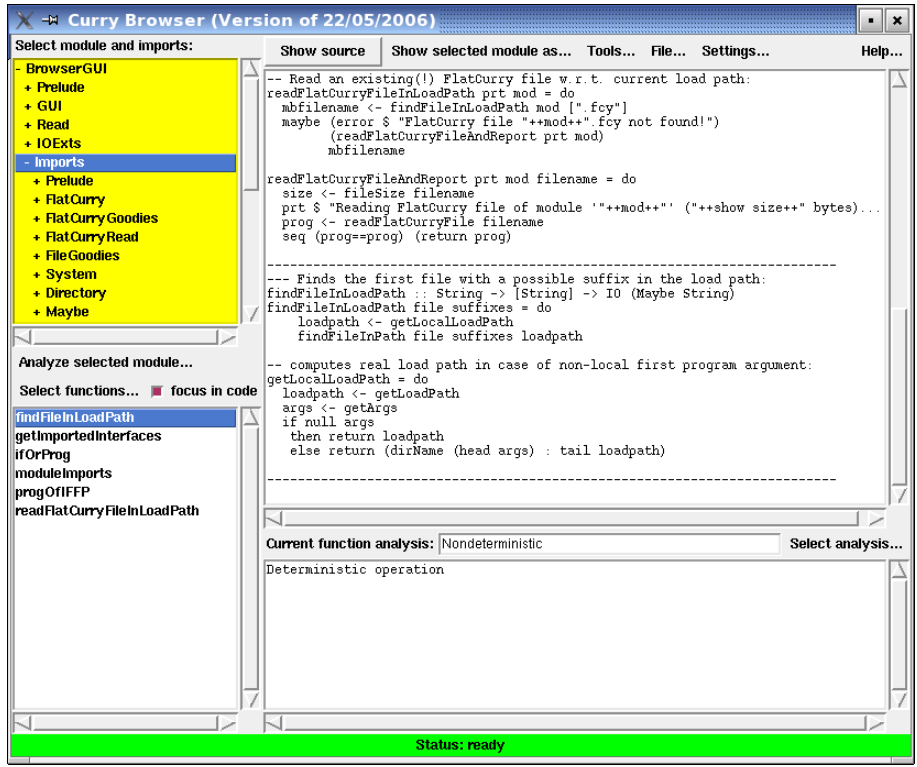,scale=0.57}
\end{center}
\caption{The main window of CurryBrowser\label{fig-main}}
\end{figure*}
To get an impression of the use of \cb, Figure~\ref{fig-main}
shows a snapshot of its use on a particular application
(here: the implementation of \cb).
The upper list box in the left column shows the modules and their imports
in order to browse through the modules of an application.
Similarly to directory browsers, the list of imported modules of a module
can be opened or closed by clicking.
After selecting a module in the list of modules, its source code,
interface, or various other formats of the module can be shown
in the main (right) text area. For instance, one can show
pretty-printed versions of the intermediate flat programs (see below)
in order to see how local function definitions are translated by lambda lifting
\cite{Johnsson85}
or pattern matching is translated into case expressions \cite{Hanus97POPL,Wadler87}.
Since Curry is a language with parametric polymorphism and type inference,
programmers often omit the type signatures when defining functions.
Therefore, one can also view (and store) the selected module as source code where
missing type signatures are added.

Below the list box for selecting modules, there is a menu
(``Analyze selected module'') to analyze all functions
of the currently selected module at once. This is useful
to spot some functions of a module that could be problematic
in some application contexts, like functions that are impure (i.e., the result
depends on the evaluation time) or partially defined (i.e.,
not evaluable on all ground terms).
If such an analysis is selected,
the names of all functions are shown in the
lower list box of the left column (the ``function list'')
with prefixes indicating the properties of the individual functions.

The function list box can be also filled with functions
via the menu ``Select functions''. For instance, all functions
or only the exported functions defined in the currently selected
module can be shown there, or all functions from different modules
that are directly or indirectly called from
a currently selected function.
This list box is central to focus on a function in the
source code of some module or to analyze some function,
i.e., showing their properties. In order to focus on a function,
it is sufficient to check the ``focus on code'' button.
To analyze an individually selected function, one can
select an analysis from the list of available program analyses
(through the menu ``Select analysis'', see also Figure~\ref{fig-mainmenu}).
In this case, the analysis results are either shown
in the text box below the main text area
or visualized by separate tools, e.g., by a graph drawing tool for
visualizing call graphs.
Some analyses are local, i.e., they need only to consider the local definition
of this function (e.g., ``Calls directly,'' ``Overlapping rules,''
``Pattern completeness'', see Section~\ref{sec-available-tools}),
where other analyses are global, i.e.,
they consider the definitions of all functions directly or indirectly called
by this function (e.g., ``Depends on,'' ``Solution complete,''
``Set-valued'').
Finally, there are a few additional tools integrated into \cb,
for instance, to visualize the import relation between all modules
as a dependency graph. These tools are available through the ``Tools'' menu.

In the next section, we review some features of Curry
in order to show some details of the implementation of \cb
in Section~\ref{sec-impl}.
The currently available analyses and tools are sketched in
Section~\ref{sec-available-tools} before we conclude in
Section~\ref{sec-concl}.

\section{Curry Programs}
\label{sec-intro-curry}

Since \cb is implemented in Curry and intended to be applied to Curry programs,
we review in this section some aspects of Curry programs
that are necessary to understand the functionality and
implementation of our programming environment.
More details about Curry's computation model and a complete
description of all language features can be found in
\cite{Hanus97POPL,Hanus06Curry}.

Curry is a declarative multi-paradigm language
combining in a seamless way features from functional,
logic, and concurrent programming
and supports programming-in-the-large with specific features
(types, modules, encapsulated search).
From a syntactic point of view, a Curry program is a functional
program extended by the possible inclusion of free (logic)
variables in conditions and right-hand sides of defining rules.
Curry has a Haskell-like syntax \cite{PeytonJones03Haskell},
i.e., (type) variables and function names usually
start with lowercase letters and the names of type and data constructors
start with an uppercase letter. The application of a function $f$
to an argument $e$ is denoted by juxtaposition (``$f~e$'').

A \emph{Curry program} consists of the definition of functions
and the data types on which the functions operate.
Functions are defined by conditional equations with constraints in the conditions.
They are evaluated lazily and can be called with partially instantiated arguments.

\begin{example}\rm
\label{ex-conc}
The following program defines the types of
Boolean values and polymorphic lists
and functions to concatenate lists and to compute the last
element of a list:
\startprog
data Bool   = True | False
data List a = []   | a : List a
\medskip
conc :: [a] -> [a] -> [a]
conc []     ys = ys
conc (x:xs) ys = x : conc xs ys
\medskip
last xs | conc\,\,ys\,\,[x] =:= xs  = x  where\,\,x,ys\,\,free
\stopprog
The data type declarations define
\code{True} and \code{False} as the Boolean constants and
\code{[]} (empty list) and \code{:} (non-empty list) as the constructors for
polymorphic lists (\code{a} is a type variable ranging over
all types and the type \ccode{List\,\,a} is usually written as \code{[a]}
for conformity with Haskell).
The (optional) type declaration (\ccode{::}) of the function \code{conc}
specifies that \code{conc} takes two lists as input and produces
an output list, where all list elements are of the same
(unspecified) type.\footnote{Curry uses curried function types
where \code{$\alpha$->$\beta$} denotes the type of all functions
mapping elements of type $\alpha$ into elements of type $\beta$.}
\end{example}

The operational semantics of Curry \cite{AlbertHanusHuchOliverVidal05,Hanus97POPL}
is a conservative extension of lazy functional programming (if free variables
do not occur in the program or the initial goal) and (concurrent) logic programming.
To describe this semantics, compile programs, or implement analyzers and
similar tools, an intermediate representation of Curry programs has been
shown to be useful.
Programs of this intermediate language, also called \emph{flat programs},
\label{sec-flat-programs}
contain a single rule for each function where the pattern matching strategy
is represented by case/or expressions.
The basic structure of flat programs is defined as follows:\footnote{%
$\ol{o_k}$ denotes a sequence of objects $o_1,\ldots,o_k$.}\\[2ex]
{\small 
$
\begin{array}{@{~~~}lcl@{\hspace*{15ex}}lcl}
P & ::= & D_1 \ldots D_m & e & ::= & v \\
D & ::= & f~v_1 \,\ldots\, v_n = e & 
  & | & c~e_1\,\ldots\, e_n  \\
&&&  & | & f~e_1 \,\ldots\, e_n  \\
p & ::= & c~v_1 \,\ldots\, v_n  &
  & | & \mathit{case}~e_0~\mathit{of}~\{\ol{p_k\to e_k}\} \\
  & & &
  & | & \mathit{fcase}~e_0~\mathit{of}~\{\ol{p_k\to e_k}\} \\
  & & &
  & | & e_1~\mathit{or}~e_2 \\
\end{array}
$}\\[2ex]
A program $P$ consists of a sequence of
function definitions $D$ with pairwise different variables in the left-hand sides.
The right-hand sides are expressions $e$ composed by variables, constructor and
function calls, case expressions, and disjunctions.
A case expression has the form
$
\mathit{(f)case}~e~\mathit{of}~
\{c_1~\ol{x_{n_1}} \to e_1,\ldots,c_k~\ol{x_{n_k}} \to e_k\}
$,
where $e$ is an expression, $c_1,\ldots,c_k$ are different 
constructors of the
type of $e$, and $e_1,\ldots, e_k$ are expressions.
The \emph{pattern variables} $\ol{x_{n_i}}$ are local
variables which occur only in the corresponding subexpression $e_i$.
The difference between $\mathit{case}$ and $\mathit{fcase}$ shows up when the
argument $e$ is a free variable:
$\mathit{case}$ suspends (which corresponds to residuation)
whereas $\mathit{fcase}$ nondeterministically binds this variable
to the pattern in a branch of the case expression
(which corresponds to narrowing).

The PAKCS implementation of Curry \cite{Hanus06PAKCS} provides a library
for meta-programming which contains the data types
for representing flat programs (i.e., the data types and functions defined
in a module)
and an I/O action for reading a module and translating
its contents into the corresponding data term.
For instance, a module of a Curry program
is represented as an expression of type
\startprog
data Prog = Prog String [String] [TypeDecl] [FuncDecl] [OpDecl]
\stopprog
where the arguments of the data constructor \code{Prog}
are the module name, the names of all imported modules,
the list of all type, function, and infix operator declarations.
Furthermore, a function declaration is represented as
\startprog
data FuncDecl = Func QName Int Visibility TypeExpr Rule
\stopprog
where the arguments are the qualified name (i.e., a pair of module and function name),
arity, visibility (\code{Public} or \code{Private}), type, and rule
(of the form \ccode{Rule $\mathit{arguments}$ $\mathit{expr}$})
of the function.
The remaining data type declarations for representing Curry programs
are similar but we omit them for brevity.

\section{Implementation}
\label{sec-impl}

\cb is implemented in Curry using libraries for GUI programming \cite{Hanus00PADL}
and meta-programming sketched above.
In order to ensure a fast startup time, only the interface files of all modules
(they have the same structure as flat programs but contain only the type
signatures of exported functions and data types) are read at the beginning.
This information is sufficient to show the import structure of all modules
in the initial main window.
Complete flat programs are only read as demanded by the analyses.

As discussed in Section~\ref{sec-overview}, \cb offers a generic interface
to integrate various analysis tools for declarative programs.
Since flat programs are an appropriate abstraction level for implementing
such tools \cite{BrasselHanus05,HanusKoj01WLPE,HanusSteiner00PPDP},
this interface is based on them.
To be more concrete, \cb provides the following type definition to
connect program analyzers (where \code{a} is the type of the analysis result):
\startprog
data FunctionAnalysis a =
     LocalAnalysis      (FuncDecl -> a)
   | LocalDataAnalysis  ([TypeDecl] -> FuncDecl -> a)
   | GlobalAnalysis     ([FuncDecl] -> [(QName,a)])
   | GlobalDataAnalysis ([TypeDecl] -> [FuncDecl] -> [(QName,a)])
\stopprog
A local analysis associates results to single function definitions
(e.g., ``Calls directly''),
a local data analysis requires in addition the type declarations
(e.g., ``Pattern completeness''), and global analyses require all defined
functions and yield lists containing for each function name (\code{QName})
an associated result. Thanks to laziness, the results for all functions
are not computed at once but only as demanded by the user.

As a simple example, consider the ``Overlapping rules'' analysis.
This analysis is intended to indicate whether a function is defined by rules
with overlapping left-hand sides.
Although this analysis is simple, it is interesting since
functions (sometimes accidentally) defined by overlapping rules
might cause nondeterministic evaluations even for ground expressions.
This property can be easily spotted in the flat representation of Curry programs
as an occurrence of a disjunction ($\mathit{or}$) in the right-hand side
of the rule defining this function. Thus, the analysis is a local one
that can be implemented as follows:
\startprog
isOverlappingFunction :: FuncDecl -> Bool
isOverlappingFunction (Func _ _ _ _ (Rule _ e)) = orInExpr e
\stopprog
where the operation \code{orInExpr} checks for occurrences of disjunctions
in an expression.

For the simple addition of new analyzers, the implementation of \cb
has a configuration module containing definitions of the currently available tools.
For instance, it contains a constant \code{functionAnalyses} of type
\startprog
functionAnalyses :: [(String, FunctionAnalysis AnalysisResult)]
\stopprog
where each element in this list consists of the name and the operation
implementing the analysis. The result type of a concrete function analysis
indicates the way the result is handled inside \cb.
Currently, this type is defined as
\startprog
data AnalysisResult = MsgResult    String
                    | ActionResult (IO ())
\stopprog
Thus, the analysis result is a string to be shown inside the \cb interface
or an I/O action to visualize the result via an external tool (e.g.,
a graph drawing tool).
Thus, to test a new analysis by integrating it into \cb,
one has to connect its implementation by adding a new element to the
list \code{functionAnalyses} and recompile the system.
Then, the \cb environment provides the analysis with the appropriate data
whenever the user selects the analysis.

\begin{figure*}[t]
\begin{center}
  \epsfig{file=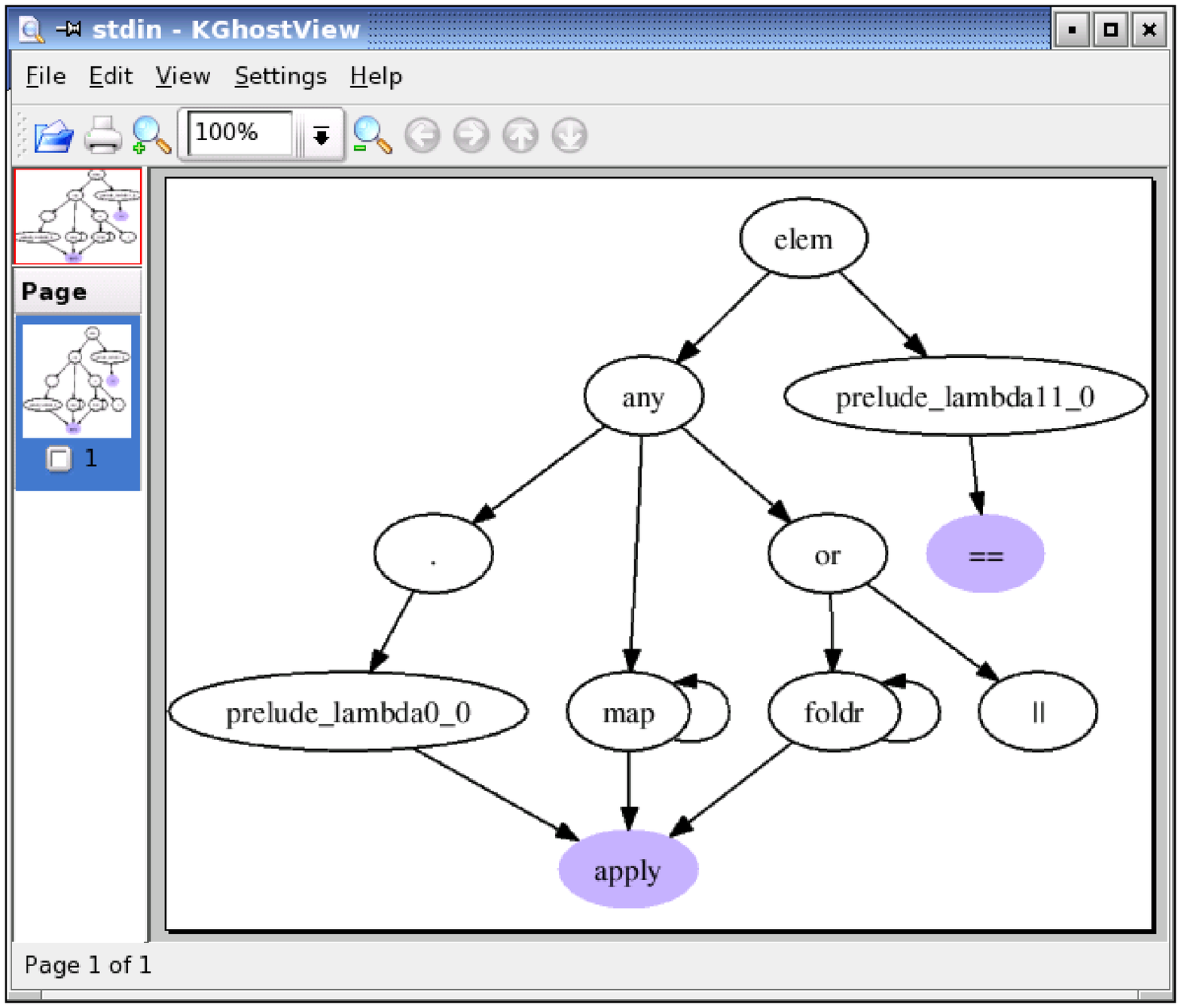,scale=0.55}
\end{center}
\caption{Visualization of the dependency graph of the prelude function \code{elem}\label{fig-depgraph}}
\end{figure*}
Note that the possible result types of analyses to be integrated into \cb
are fixed\footnote{The possible analysis results might
be extended in future version of \cb. For instance, one could map
complex analysis results into a source code transformation that is shown
in the code widget.}
since the implementation needs to know what to do with the analysis
result. Therefore, the current implementation supports
\begin{itemize}
\item string results that are shown inside the main window, or
\item I/O actions that calls some external tool for visualization.
\end{itemize}
For instance, to show the dependency graph of a function,
the corresponding global analysis computes a graph structure and calls an
external program, the DOT graph drawing tool\footnote{{\tt http://www.graphviz.org/}},
to visualize this graph structure
(see Figure~\ref{fig-depgraph} for an example).
Since the main GUI of \cb is executed in the I/O monad,
the event handlers that implement reactions to user events are
I/O actions \cite{Hanus00PADL}. Thus, analyses with a result type \code{IO\,()}
can be executed by the event handlers responsible for analyzing programs.
In order to avoid the crash of the \cb environment if some analyses fails
with run-time errors, the execution of an analyses is wrapped into an
exception handler.

The restriction to a fixed set of analysis result types requires the transformation
of arbitrary program analyses when they are integrated into \cb.
For instance, the ``Overlapping rules'' analysis sketched above
delivers Boolean results that must be converted into appropriate strings
shown to the user. For this purpose, one can define a simple
conversion operation to show the result of the overlapping analysis:
\startprog
showOverlap :: Bool -> String
showOverlap True  = "Overlapping"
showOverlap False = "Not Overlapping"
\stopprog
In order to support a simple conversion of arbitrary analyses
into the analyses with string results as required by the interface of \cb,
the implementation of \cb contains the following conversion operation:
\startprog
showWithMsg :: FunctionAnalysis a
               -> (a->String)
               -> FunctionAnalysis AnalysisResult
\medskip
showWithMsg (LocalAnalysis ana) showresult =
  LocalAnalysis (\ttbs{}f -> MsgResult (showresult\,(ana f)))
showWithMsg (LocalDataAnalysis ana) showresult =
  LocalDataAnalysis (\ttbs{}tds f -> MsgResult (showresult\,(ana tds f)))
showWithMsg (GlobalAnalysis ana) \ldots
showWithMsg (GlobalDataAnalysis ana) \ldots
\stopprog
Based on this conversion operation (which is also defined as an infix operator),
it is easy to integrate an analysis like \code{isOverlappingFunction}
into \cb by adding an element to the configuration list \code{functionAnalyses}
as follows:
\startprog
functionAnalyses =
 [\ldots,
  ("Overlapping rules",
   LocalAnalysis isOverlappingFunction
                           `showWithMsg` showOverlap),
  \ldots]
\stopprog
Similarly to the list constant \code{functionAnalyses},
the configuration module of \cb contains two further list constants that
specify the available analyses:
\begin{itemize}
\item
a list constant \code{allFunctionAnalyses} that contains
analyses that are applied to \emph{all} functions of the selected module
(e.g., applying the analysis ``Pattern completeness'' to all functions
of a module is useful to spot those functions with incomplete pattern definitions):
the results of these analyses are shown as prefixes in the column showing the
list of all functions of the currently selected module;
\item
a list constant \code{moduleAnalyses} that contains analyses
that are applied to complete modules (e.g., to generate the interface,
the flat representation of a module, or a source code representation where
missing type signatures are added); similarly to a function analysis,
the result of a module analysis is either a (program) text to be shown
in the main window or an I/O action that visualizes the result by an external tool.
\end{itemize}
Thus, it is fairly easy to integrate existing tools (implemented in Curry)
into \cb.

\begin{figure*}[t]
\begin{center}
  \epsfig{file=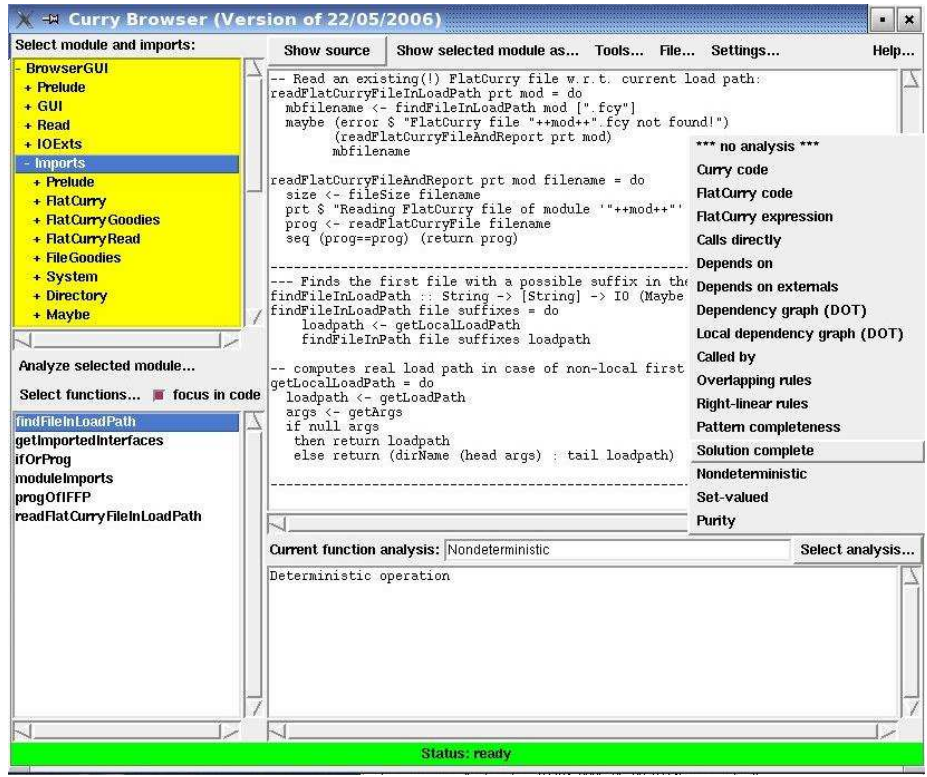,scale=0.42}
\end{center}
\caption{Selection of a function analysis\label{fig-mainmenu}}
\end{figure*}

\section{Available Analyses and Tools}
\label{sec-available-tools}

This section shortly surveys the analyses and tools that are
currently available in \cb (see Figure~\ref{fig-mainmenu}).
Due to the simple integration of further analyses and tools,
this set is likely to be extended in future releases of \cb.

In the flat representation of Curry (see Section~\ref{sec-flat-programs}),
pattern matching is made
explicit by case expressions and disjunctions, and local definitions
are ``globalized'' by lambda lifting \cite{Johnsson85}.
Although it is possible to deal with local definitions at run time
(e.g., \cite{AntoyJohnson04PPDP}),
lambda lifting is useful to simplify the operational model
\cite{AlbertHanusHuchOliverVidal05} of Curry programs
and, thus, often used in implementations of functional logic languages
(e.g., \cite{AntoyHanus00FROCOS,AntoyHanusMasseySteiner01PPDP,LuxKuchen99}.
Thus, it is sometimes interesting to show the effect of these transformations
performed by the front end of a Curry implementation.
For this purpose, \cb can show the flat representation of each function
or module as well as a source-like representation where case expressions
and disjunctions are translated back into pattern-based definitions
but local definitions are kept globalized to get an impression
of the program usually passed to a back end.
Furthermore, the following analyses are available for individual functions:
\begin{description}
\item[\code{Calls directly} {\rm(local analysis):}]
Shows all functions that are directly called by this function.

\item[\code{Depends on} {\rm(global analysis):}]
Shows all functions that might be directly or indirectly called in the rules
defining this function.

\item[\code{Dependency graph} {\rm(global analysis):}]
Shows the dependency graph of the selected function.
This is a combination as well as a graphical visualization of
\code{Calls directly} and \code{Depends on}, i.e.,
an arc is drawn from each function symbol
to all functions directly called in the rules defining this function
and all reachable function nodes are included in the graph
(see Figure~\ref{fig-depgraph} for an example).

\item[\code{Local dependency graph} {\rm(global analysis):}]
Shows the dependency graph of the selected function restricted to all rules
occurring in the current module. This is useful for complex functions,
e.g., depending on other non-trivial library functions,
where the complete dependency graph becomes unreadable due to its size.

\item[\code{Called by} {\rm(global analysis):}]
Shows the list of all functions in the current module
that call this function in their defining rules.

\item[\code{Overlapping rules} {\rm(local analysis):}]
Shows whether the function is defined by overlapping rules
(which might cause nondeterministic evaluations even for ground expressions).
This is interesting for logic programming but might be also useful
for purely functional programs.

\item[\code{Right-linear rules} {\rm(local analysis):}]
Shows whether the function is defined by right-linear rules,
i.e., rules where each variable has at most one occurrence in the right-hand side.

\item[\code{Right-linearity} {\rm(global analysis):}]
Shows whether the function is defined by right-linear rules
and depends only on functions defined by right-linear rules.
This information is useful for some program optimizations
(e.g., \cite{AntoyHanus05LOPSTR}).

\item[\code{Pattern completeness} {\rm(local data analysis):}]
Shows whether the pattern matching is exhaustive, i.e.,
if the function is reducible
on any combination of (well-typed) ground constructor argument terms.

\item[\code{Totally defined} {\rm(global data analysis):}]
Shows whether the function is totally defined, i.e.,
if the function evaluates to a value
for any combination of (well-typed) ground constructor argument terms.
This is the case if it directly or indirectly depends only on operations
that are pattern complete.

\item[\code{Solution completeness} {\rm(global analysis):}]
Shows whether the function is operationally complete, i.e.,
if it is ensured that the execution of the function does
not suspend for any arguments.

\item[\code{Nondeterministic} {\rm(global analysis):}]
Shows whether the function is possibly nondeterministic, i.e.,
if it directly or indirectly depends on an operation
defined by overlapping rules so that it might deliver
(nondeterministically) two values for the same ground constructor
arguments.

\item[\code{Set-valued} {\rm(global analysis):}]
Shows whether the function is possibly set-valued, i.e.,
if it directly or indirectly depends on an operation
defined by overlapping rules or rules containing extra variables
(variables occurring in the right-hand side but not in the left-hand side)
so that an application to some ground constructor arguments is equal to
a set of more than one value. For instance, the prelude function
\code{unknown} defined by
\startprog
unknown = x  where x free
\stopprog
is set-valued but not nondeterministic: the evaluation of \code{unknown}
is deterministic (and yields a fresh logic variable) but, from a semantical
point of view, it is set-valued since it denotes any possible value.

\item[\code{Purity} {\rm(global analysis):}]
Shows whether the function is pure (referentially transparent), i.e.,
if it is ensured that it delivers always the same values for the same ground constructor
arguments at each time and all schedulings of the evaluation.
This might not be the case if committed choice or sending via ports
is executed during its evaluation \cite{Hanus99PPDP}.
\end{description}
Finally, there are also useful tools to process complete modules or collections
of modules.
For instance, beyond showing the source code, interface, and flat representation
of a module, one can also show a version of the source code where
type signatures are added to functions where the programmer has omitted them.

\begin{figure*}[t]
\begin{center}
  \epsfig{file=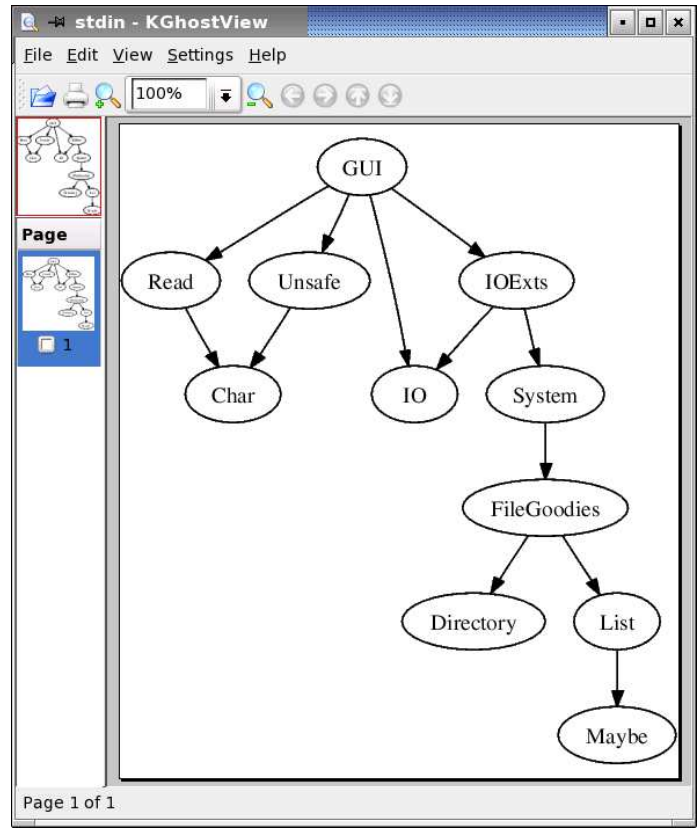,scale=0.6}
\end{center}
\caption{Module dependency graph of the library \code{GUI} used in the
   \cb implementation (visualized by \cb)\label{fig-importgraph}}
\end{figure*}
To get more information about the import structure between modules,
one can show all modules directly imported by the current one
together with their exported functions that are accessed in the current module.
This is useful to spot superfluously imported modules.
Finally, one can also visualize the entire import relation between all modules
of the currently loaded application as a module dependency graph
(see Figure~\ref{fig-importgraph} for an example).

\section{Conclusions}
\label{sec-concl}

We have presented \cb, a generic analysis environment for Curry programs.
\cb supports browsing through the modules of an application
and offers a wide range of analysis tools in an integrated manner.
The currently available set of analyses and tools can be easily
extended due to the generic interface offered by the implementation of \cb.

The mostly related system (and, in some sense, its predecessor)
is CIDER \cite{HanusKoj01WLPE}, an environment to analyze single Curry modules.
In contrast to CIDER, \cb can be applied to complete applications
consisting of several modules and supports the browsing through the module structure.
Furthermore, \cb provides a better structure to integrate analyzers:
CIDER assumes that every analysis takes the complete program as input,
whereas \cb distinguishes between different kinds of analyses
(local, global, data-dependent) and provides them with the appropriate
information from the modules and functions selected by the user.

Another related system is ${\cal IDE}$ \cite{DiosCastroGonzalezMoreno00},
a graphical development environment for the functional logic
languages Toy and Curry.
${\cal IDE}$ supports the writing of programs in a standard
text editor window and the compilation and execution of programs.
However, ${\cal IDE}$ does not offer further tools, e.g., for program analysis.
This is in contrast to
the Ciao Preprocessor (CiaoPP) \cite{HermenegildoPueblaBuenoLopez05},
a tool integrating sophisticated program analyses
with validation and transformation methods for logic programs.
The emphasis of CiaoPP is the use of program analyses
to manipulate logic programs rather than a graphical programming environment
supporting an easy way to browse through programs, as in \cb.

\cb is completely implemented in Curry.
The advanced programming techniques offered by Curry
(e.g., higher-order functions, demand-driven evaluation, meta-programming,
high-level abstractions with logic variables for GUI programming
\cite{Hanus00PADL})
has supported the fast and maintainable implementation of this environment.
The size of the complete implementation of \cb is
approximately 1400 lines of Curry code. This includes
the implementation of the graphical user interface
and all currently available analyses and tools.
In addition, the total size of all imported system libraries is approximately
2500 lines of Curry code.
These numbers provide an indication of the advantages
obtained by the use of declarative high-level programming languages
for the implementation of complex systems.
The implementation of \cb is freely available with the latest distribution
of PAKCS \cite{Hanus06PAKCS} where it has been integrated for easy use.

For future work, it is interesting to integrate further tools
into \cb, like tools for program transformation
(e.g., partial evaluation \cite{AlbertHanusVidal02JFLP},
refactoring \cite{Thompson05}),
program observation \cite{BrasselChitilHanusHuch04PADL},
tracing \cite{BrasselHanusHuchVidal04}, in order to obtain
a comprehensive programming environment.


\end{document}